\documentclass[twoside]{ilcws10}
\usepackage[utf8]{inputenc}
\usepackage{graphicx,epsfig,color}
\usepackage{wrapfig,rotating}
\usepackage{amssymb,amsmath,array}
\usepackage{subfig}
\usepackage{multicol}

\newcommand{\physicsList}[1]{\textsc{\texttt{#1}}}
\newcommand{\qgspBert}{\physicsList{QGSP\_BERT}}
\newcommand{\qgspBertTrv}{\physicsList{QGSP\_BERT\_TRV}}
\newcommand{\qgsBic}{\physicsList{QGS\_BIC}}
\newcommand{\lhep}{\physicsList{LHEP}}
\newcommand{\ftfBic}{\physicsList{FTF\_BIC}}
\newcommand{\ftfpBert}{\physicsList{FTFP\_BERT}}

\begin{document}

\title{
Track Segments within Hadronic Showers using the CALICE AHCal}
\author{Lars Weuste$^{1,2}$
\vspace{.3cm}\\
1- Max-Planck-Institut f\"{u}r Physik, Munich, Germany \\
2- Excellence Cluster 'Universe', TU M\"{u}nchen, Garching, Germany
} 

\maketitle

\begin{abstract}
Using the high granular CALICE analog hadron calorimeter (AHCal) a tracking
algorithm capable of identifying MIP-like tracks within hadronic showers is
presented. Such an algorithm provides excellent tools for detector calibration
and for studies of the substructure of hadronic showers. 
The properties of the identified tracks are used as observables for
a Monte-Carlo to data comparison.
\end{abstract}

\section{The CALICE analog hadron calorimeter}
\label{sec:CALICE}
The CALICE (\emph{\textbf{Cal}}orimeter for a \emph{\textbf{Li}}near
\emph{\textbf{C}}ollider \emph{\textbf{E}}xperiment) collaboration performed
tests of new calorimeter technologies for the planned International Linear
Collider (ILC). Several prototypes were investigated during the testbeam phase at
DESY (2006), CERN (2006 \& 2007) and FNAL (2008 \& 2009).

The analog hadron calorimeter (AHCal, \cite{AHCal}) has been successfully used in all five
test-beams. It is a sampling calorimeter with a size of $\approx 1\;m^3$
with 38 layers using 2 cm of steel as absorbing material,
resulting in a depth of approximately $5.3\;\lambda_I$
($\approx4.5\;\lambda_I$ for pions). The active layers have a very granular
structure, using scintillator tiles from $3\times3\;cm^2$ up to
$12\times12\;cm^2$. Each scintillator tile is read out with a Silicon
Photomultiplier.

\begin{wrapfigure}{r}{0.5\columnwidth}
	\centerline{\includegraphics[width=0.45\columnwidth]{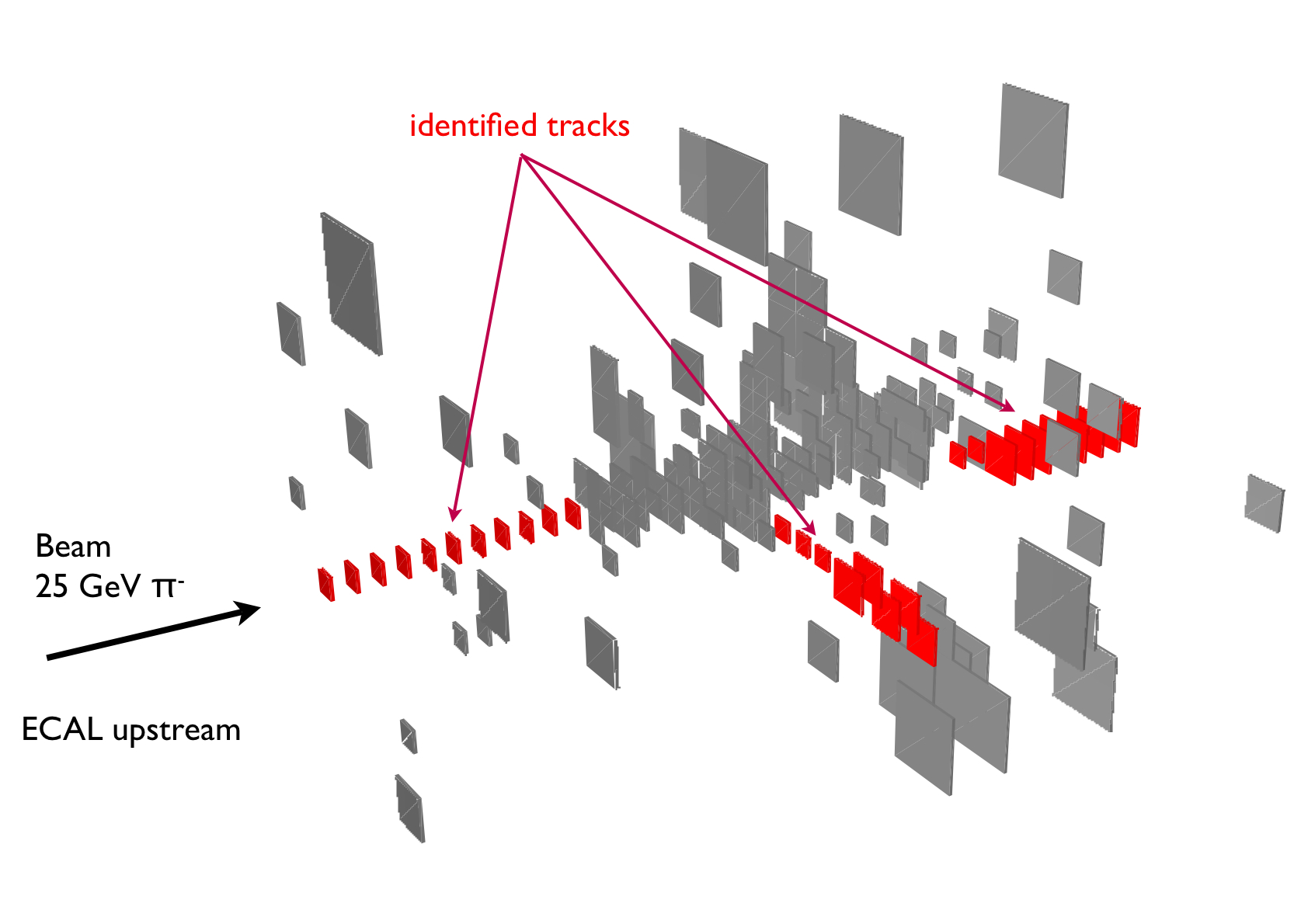}}
	\caption{Typical hadronic shower response/development in
  CALICE AHCal for a $25\;$GeV $\pi^-$ run. 
  }
	\label{fig:TypicalEvent}
\end{wrapfigure}
This granular structure allows for identification of tracks generated by Minimum
Ionizing Particles (MIP) passing through (parts of) the detector.
This capability is an excellent tool for detector calibration\cite{detCalib} and
provides the possibility for detailed studies of the shower substructure.
This article
presents an algorithm for the identification of MIP-like track segments within
hadronic showers (see Section \ref{sec:tracking}).

A few examples of identified MIP tracks are displayed in Figure
\ref{fig:TypicalEvent}. As one can see, the tracks provide information on the 3D
substructure of hadronic showers. This information is used for confronting
Monte Carlo simulations with data (Section \ref{sec:mcdata}).

\section{The tracking algorithm}
\label{sec:tracking}

\begin{figure}[h!tp]
\begin{center} 
	\subfloat[Illustration of isolation criterion]{%
	\includegraphics[height=5.5cm]{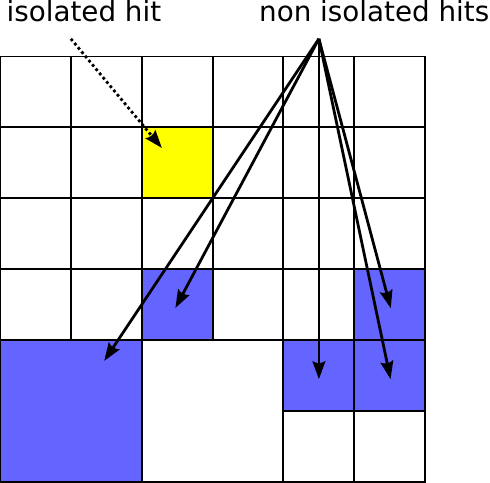}%
	\label{fig:IsolationCriterion}%
	}
 	\hspace{1cm}
	\subfloat[1D Illustration of algorithm]{%
	\includegraphics[height=5.5cm]{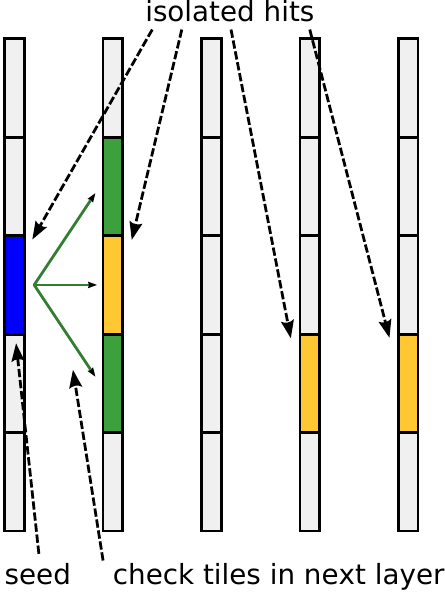}%
	\label{fig:trackingAlgorithm}%
	}
	
 	\caption{Illustration of isolation criterion and algorithm.}
 \end{center}
\end{figure}
To provide a clean sample of tracks by minimum ionizing particles in the
environment of a hadronic shower, which includes regions of high local particle
density, specific selections criteria for the calorimeter cells used by the
tracking algorithm are necessary. A sample of the detector cells very likely
being traversed only by a single particle is selected by imposing an isolation
criterion, illustrated in Figure \ref{fig:IsolationCriterion}.
A hit is called isolated if there is no direct 
neighbour cell with a hit within the same layer. 
The tracking algorithm will only use hits fulfilling this criterion.

As there is no magnetic field in the CALICE experiment, the presented algorithm
was optimized for finding straight tracks.

\subsection{Track Finding}
\label{sec:tracking:algorithm}
The tracking algorithm used for identification of single particle tracks is a
local search method working layer by layer. A track is followed from the
beginning to the end.

The algorithm has the following steps:
\begin{enumerate}
  \item Search for isolated hits in all layers using the introduced isolation
  criterion. A cell is called hit if the energy deposited exceeds $0.4\;$MIP. A 
  MIP is defined as the most probable energy deposition of a single passing minimum 
  ionizing particle.
  \item Use the isolated hits as start point (seed), starting from the first
  layer in beam direction. For each of these points:
  \begin{enumerate}
	  \item Increment by one layer in beam direction and search for unused isolated
	  hits that can continue the current track. The search window is $\pm 1$ cells
	  in $x$ and $y$ (when $z$ is beam axis), resulting in 9 checked cells. For a
	  1D illustration see Figure \ref{fig:trackingAlgorithm}. 
	  \item Repeat last step until the end of the detector is reached or no
	  continuation hit can be found. The algorithm allows for gaps, i.e. maximum
	  one consecutive layer with no hit.
  \end{enumerate}
  \item Once the track is completed, start over with the remaining isolated hits
  in the detector to find more tracks.
\end{enumerate}
As the algorithm works layer-wise for each track there is a maximum of one
hit per layer. With the isolation criterion it is then very likely that the
found track was generated by a single particle.

\section{Monte-Carlo to Data comparison}
\label{sec:mcdata}
The characteristics of MIP-like tracks reflects the spatial structure of
hadronic showers. By comparing
to simulation the shower modeling provided by the different physics lists can be
validated. Here, the two following parameters were chosen: track angle and track length.

The comparison was done using these physics lists from Mokka/Geant4\cite{Geant4IEEE}:\\
\qgspBert, \qgspBertTrv, \qgsBic, \lhep, \ftfBic, \ftfpBert

The comparison of the actual distribution is done for $\pi^-$ data at an energy
of $25\;$GeV. For energies in the range from $10\;$GeV up to
$80\;$GeV the mean value was used for comparison.

\subsection{Track angle}
\label{sec:MCData:TrackAngle}
The track angle is sensitive to the scattering angle of secondary
particles created in hadronic showers and hence the shower structure.

\begin{figure}[h!t]
\begin{center}
	\subfloat[track angle distribution for 25GeV - normalized to 1.]{%
		\includegraphics[width=.5\linewidth]{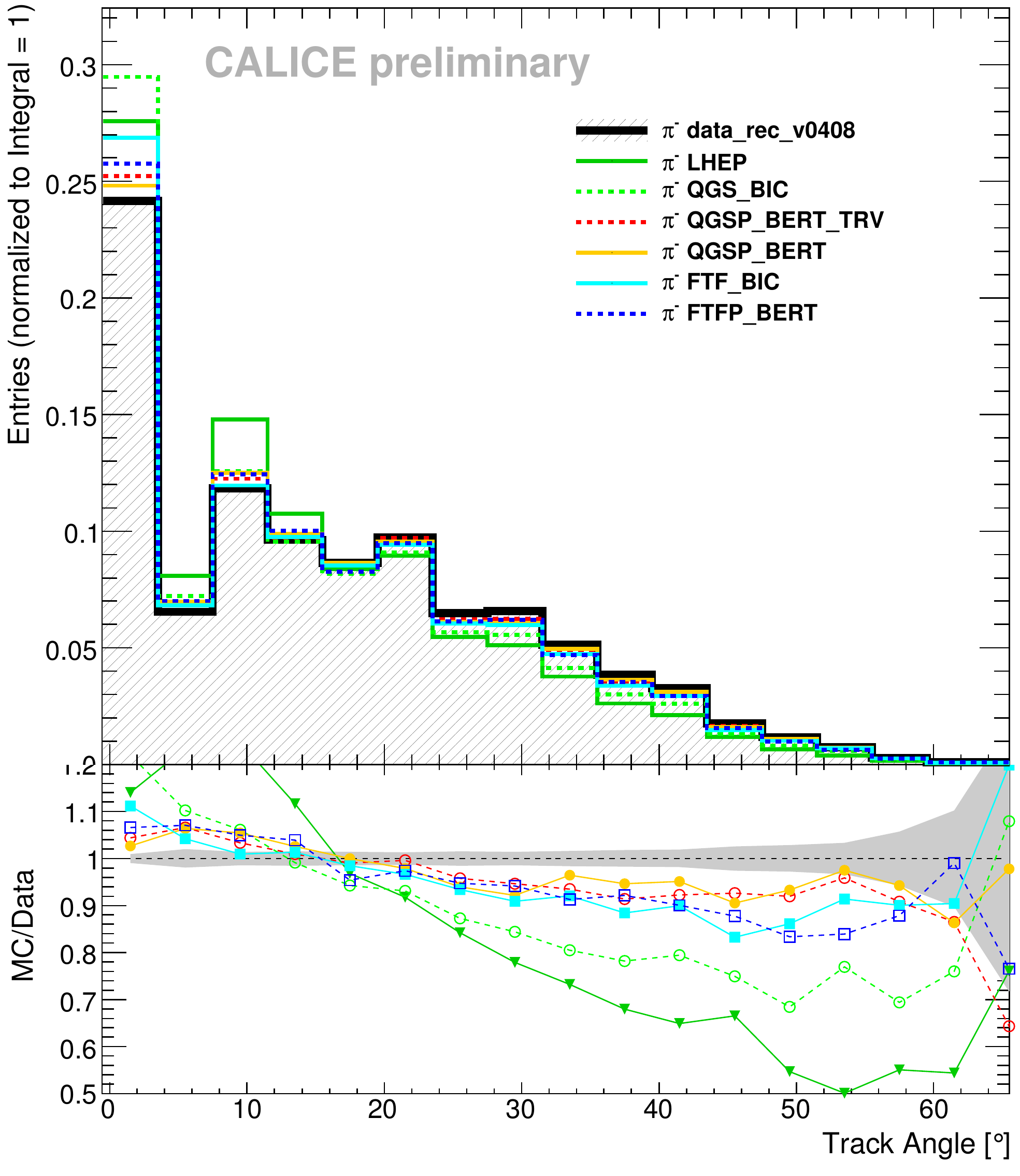}%
		\label{fig:MCData:25GeV:trackAngle}%
	}%
	\subfloat[Average track angle for all energies.]{%
		\includegraphics[width=.5\linewidth]{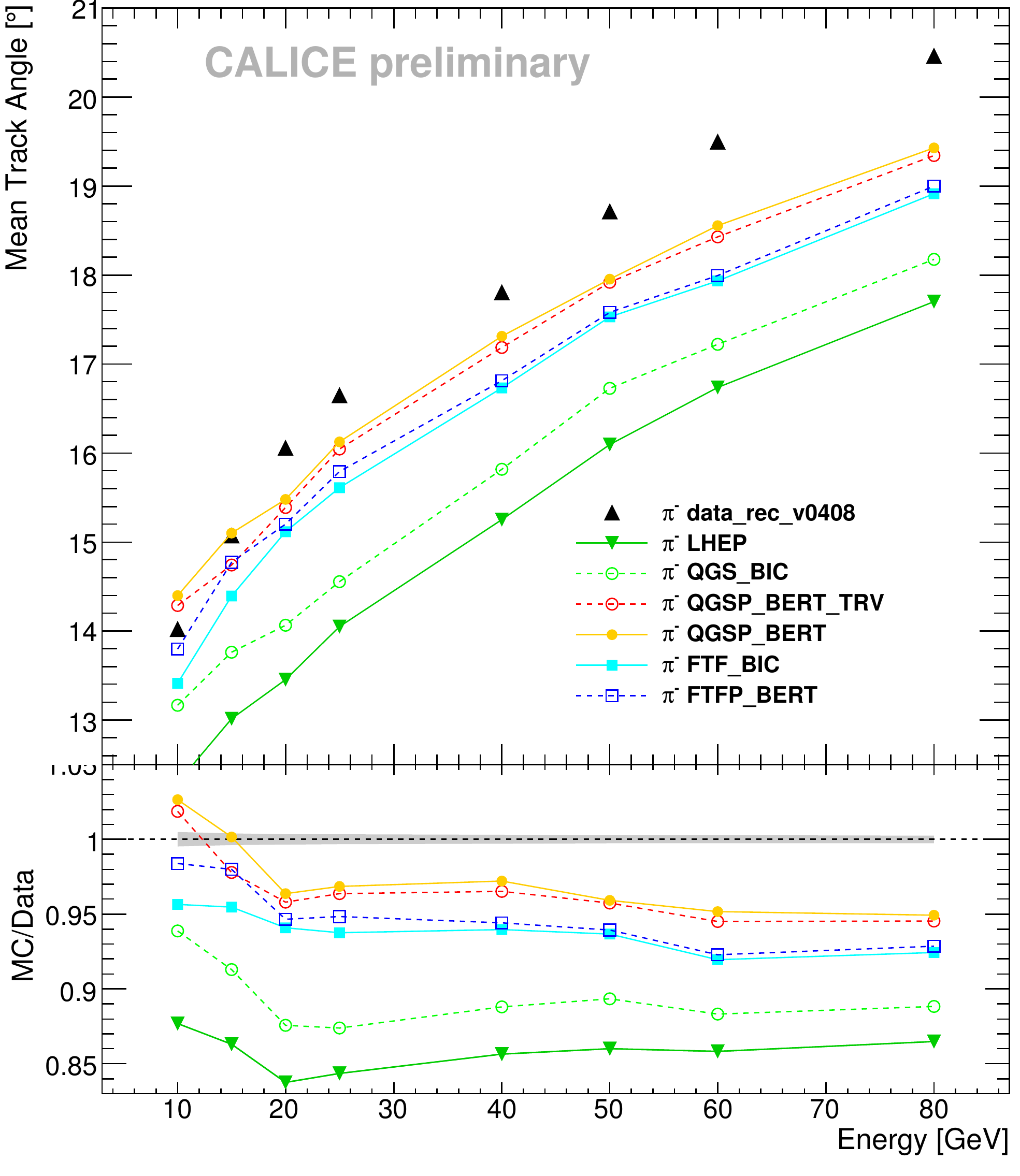}%
		\label{fig:MCData:all:trackAngle}%
	}%

  \caption[MC - data comparison: track angle]{Data - Monte Carlo comparison:
  track angle for different energies. The grey area gives the size of the
  statistical error for \lhep.}
  \label{fig:MCData:trackAngle}
\end{center}
\end{figure}

The comparison for the angle distribution of the tracks is shown in Figure
\ref{fig:MCData:25GeV:trackAngle} for $25\;$GeV. The distribution shows many
tracks at angles $\theta$ lower than $5^\circ$ ($\approx 20\;\%$
of all tracks). As in every event the incident particle is not inclined, the
initial track of the incoming particle is major contributor to those at low
angles. If we go to higher angles, the number of tracks found decreases until it
vanishes for $\theta>65^\circ$. This is not necessarily due to the shower shape
but the incapability of the used algorithm to identify tracks at very large
angles.

All physics lists considered with the exception of \lhep~and ~\qgsBic~provide
an angular distribution of tracks that is similar to the one observed in
data. The tracks produced by \lhep~and \qgsBic~have too low average angles.

This can be seen as well when comparing the mean value for the complete energy
range shown in Figure \ref{fig:MCData:all:trackAngle}. \lhep~is significantly below
the data over the whole energy range, while \qgsBic~performs better but is
still too low. The remaining physics lists are again quite close together and
reproduce the result from data better, while they all tend to produce tracks at
lower angles than testbeam data, especially for higher energies.

\subsection{Track length}
\label{sec:MCData:TrackLength}
\begin{figure}[h!tp]
\begin{center}
	\subfloat[track length distribution for 25GeV - normalized to number of events.]{%
		\includegraphics[width=.45\linewidth]{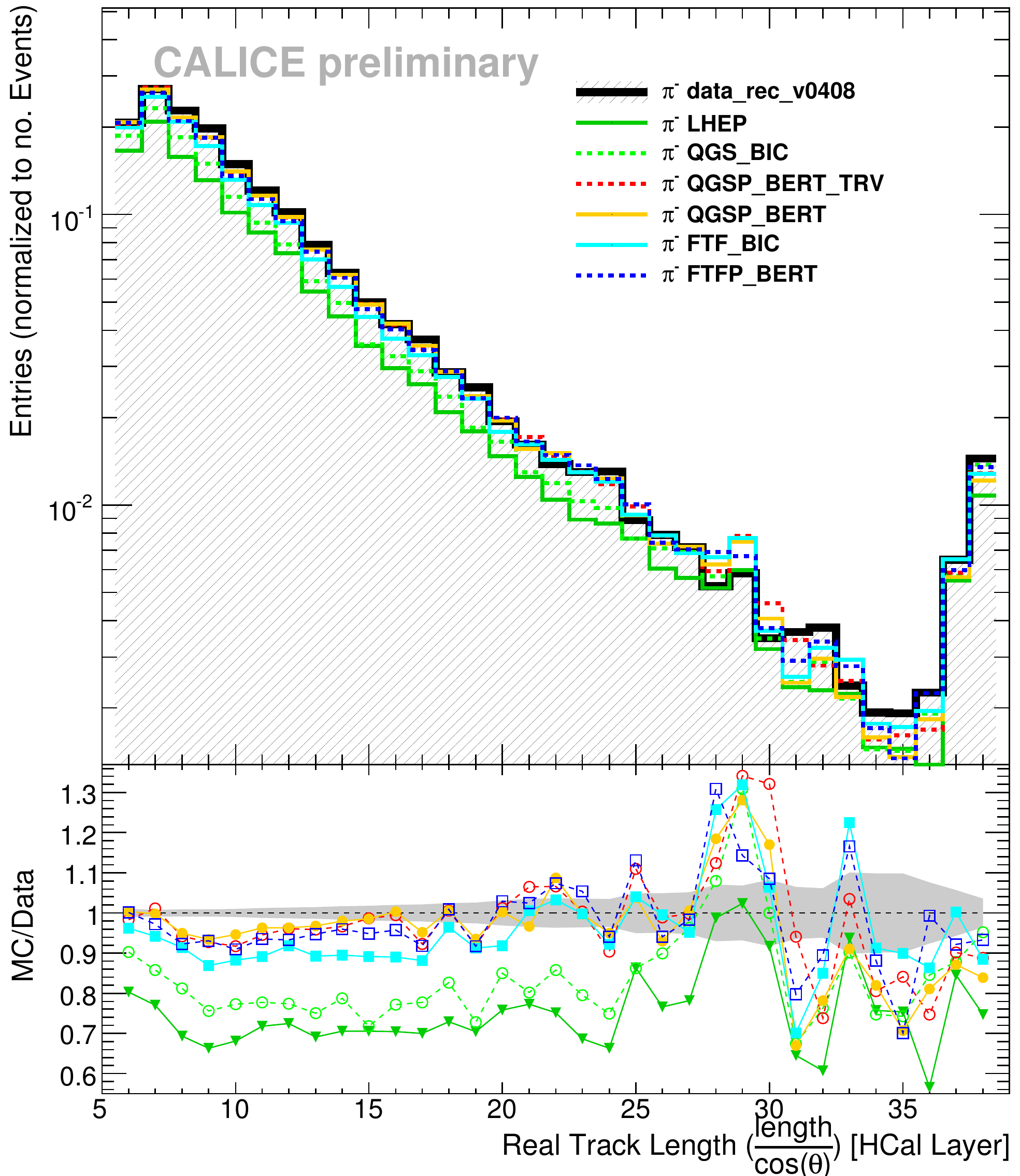}%
		\label{fig:MCData:25GeV:trackLength}%
	}%
	\subfloat[Average track length.]{%
		\includegraphics[width=.45\linewidth]{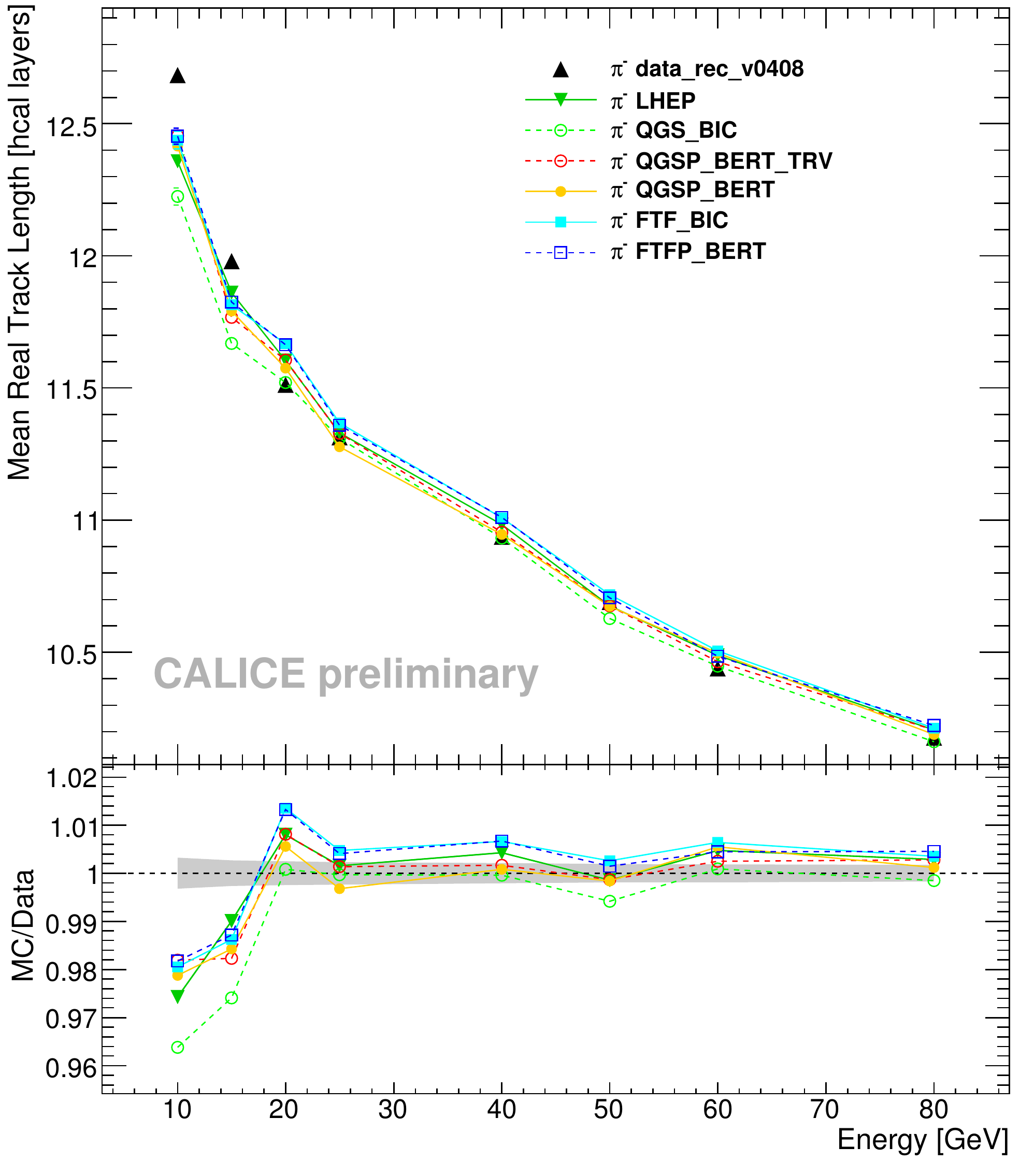}%
		\label{fig:MCData:all:trackLength}%
	}%
	\\
	\subfloat[track length distribution, starting layer 1 or 2.]{%
		\includegraphics[width=.45\linewidth]{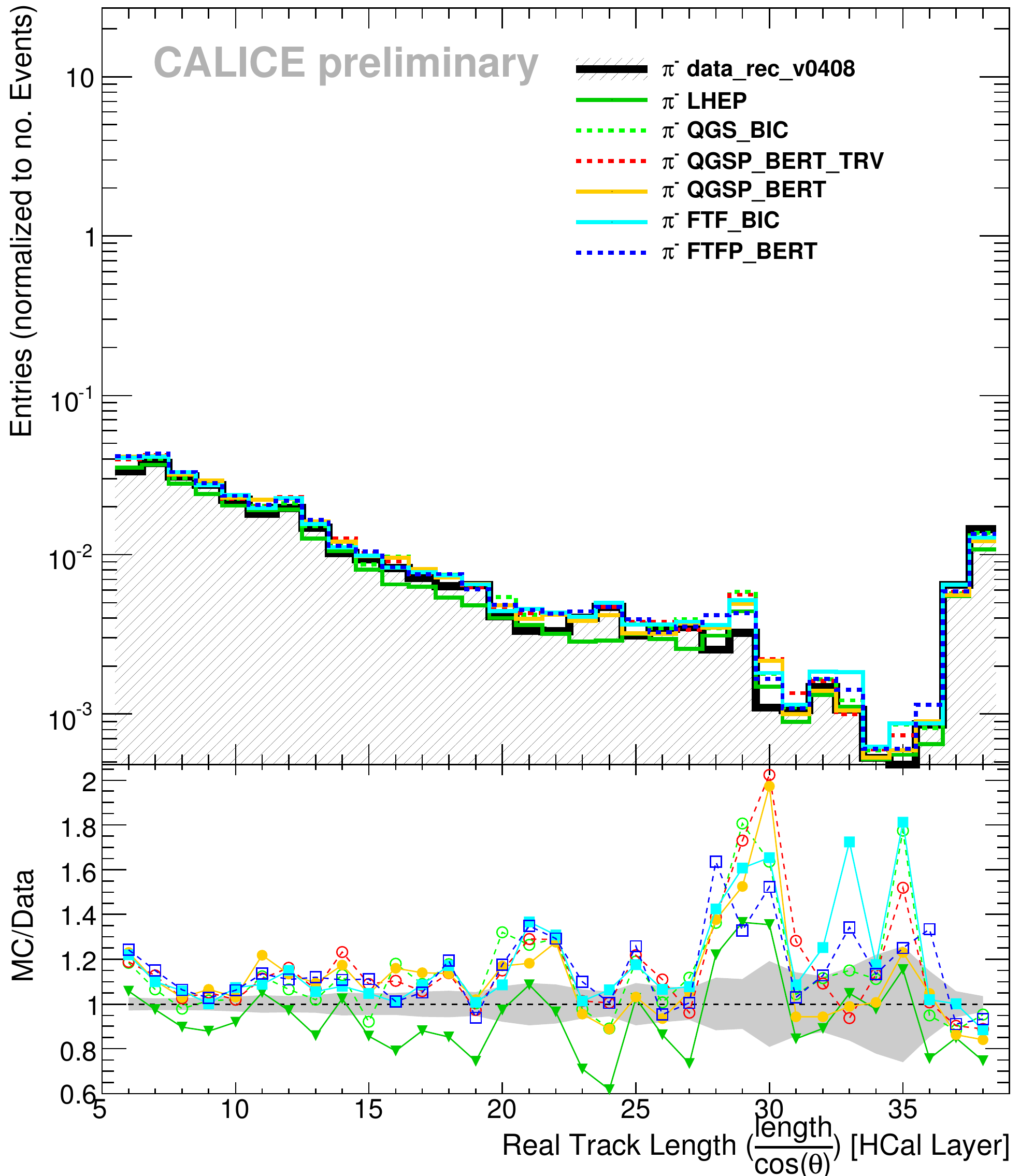}%
		\label{fig:MCData:25GeV:trackLengthFirstLayer}%
	}%
	\subfloat[track length distribution, starting layer $\geq$ 3]{%
		\includegraphics[width=.45\linewidth]{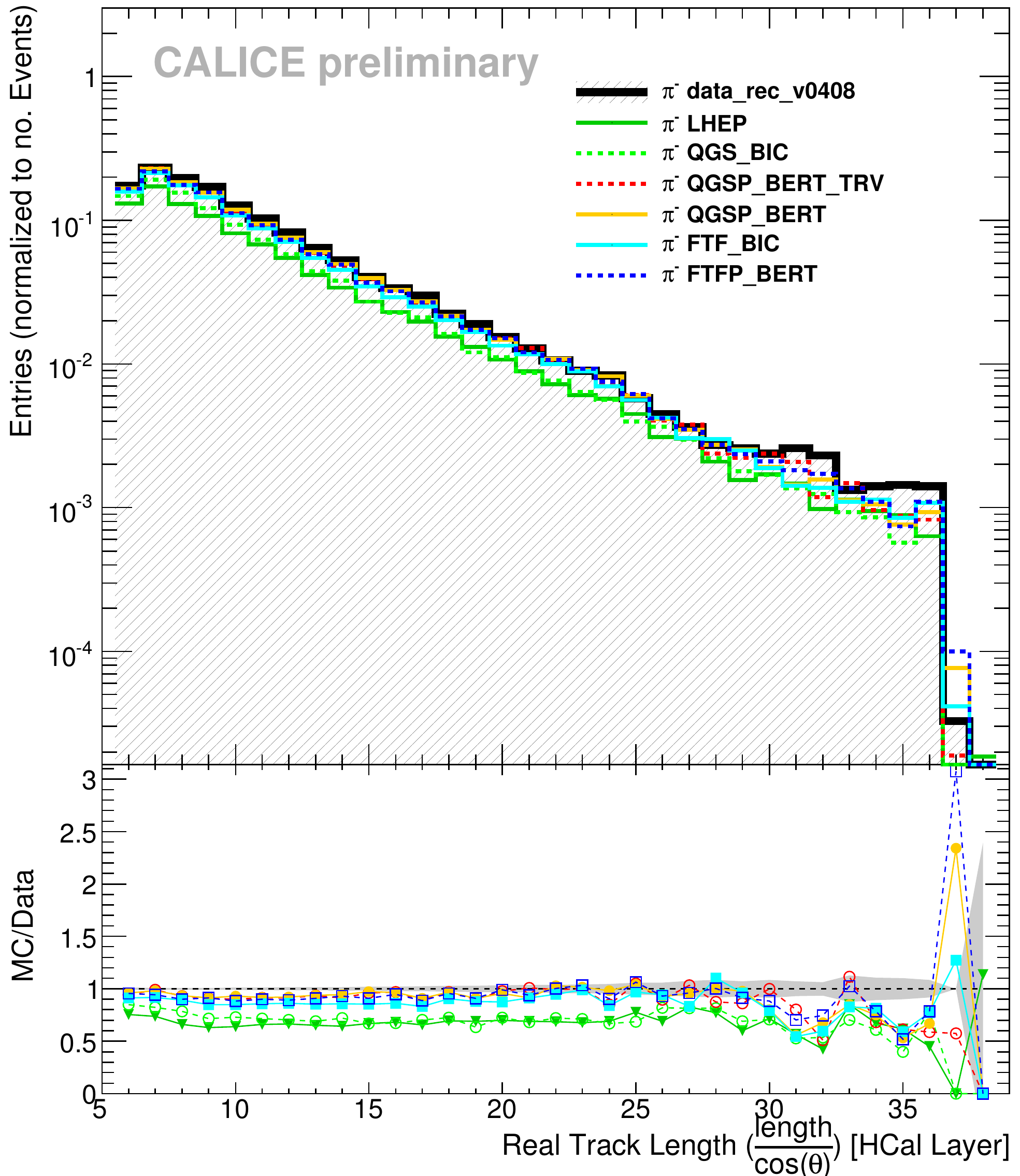}%
		\label{fig:MCData:25GeV:trackLengthAfterFirstLayer}%
	}%

  \caption[MC - data comparison: track length]{Data - Monte Carlo comparison:
   track length. \ref{fig:MCData:25GeV:trackLength} shows the full histogram for
   a $25\;$GeV run (normalized to the number of events).
    The lower plots show a decomposition of
   \ref{fig:MCData:25GeV:trackLength}, with
   \ref{fig:MCData:25GeV:trackLengthFirstLayer} showing tracks starting in layer
   1 or 2 and \ref{fig:MCData:25GeV:trackLengthAfterFirstLayer} showing all
   tracks starting in layer 3 or later. Both plots are normalized to number of
   events. Fig. \ref{fig:MCData:all:trackLength} shows the evolution of the mean
   value over the entire energy range. The gray area in all ratio plots indicates
   the size of the statistical error taken for the \lhep~physics list.}
  \label{fig:MCData:trackLength}
\end{center}
\end{figure}

The track length gives the distance a particle travels before participating in a
hadronic interaction, in units of AHCal layers:
\begin{equation}
	\textrm{real track length} = \frac{\textrm{\# layers passed}}{\cos{\theta}} 
\end{equation}
The slope of the distribution is sensitive to high energy cross sections, especially
for secondary particles created in or after the first hadronic interaction. The
results of the comparison can be seen in Fig. \ref{fig:MCData:trackLength}.

For the mean value for all energies (Figure \ref{fig:MCData:all:trackLength}) the
physics lists all show good agreement with the testbeam data. Especially for
energies higher than $20\;$GeV the difference between simulation and test
beam data is at the order of 1\%, with a slight tendency towards longer tracks in
the simulation. Noise hits can influence the isolation criterion which can lead
to tracks being aborted early. If the amount of noise is not added correctly
during the digitization phase of the simulated data, this will lead to a constant
offset between all physics lists and the testbeam data.

The track length for the $25\;$GeV run in Figure
\ref{fig:MCData:25GeV:trackLength} shows the exponential fall-off expected from
the hadronic interaction of particles passing through matter. The peak at
$38\;$layers is coming from punch through hadrons and from muons coming from
$\pi^- \to \mu^-\overline{\nu}_\mu$ decays. the amount of muons and punch-through
pions is well reproduced by the simulation.

To be able to study tracks coming from the incoming beam particle (``primary''
tracks) and tracks being created in hadronic interaction (``secondary'' tracks)
individually, the histogram from Figure \ref{fig:MCData:25GeV:trackLength} is
split into two: For the primary tracks a histogram containing only tracks
starting in layer 1 or 2 (Figure \ref{fig:MCData:25GeV:trackLengthFirstLayer})
and for the secondary tracks a histogram containing tracks starting in layer 3 or
later.

The histogram containing the secondary tracks can be seen in Figure
\ref{fig:MCData:25GeV:trackLengthAfterFirstLayer}. Those tracks are created
mainly by secondary particles coming from the shower core and hence is sensitive
to the correct modeling of high energy cross sections within the physics list.
Here the difference between testbeam data and simulation is below 10\%, with the
simulation producing less shorter tracks, but on average all physics lists
produce tracks that are longer than the ones from testbeam data, which can be
seen as well in Figure \ref{fig:MCData:all:trackLength}. The fluctuations at the
end are due to insufficient statistics (compare with statistical error indicated
by the gray area, here shown for the \lhep~physics list). The exponential decrease
is reproduced well by all physics lists, demonstrating a good description of the
cross sections in the models.

The punch-through particles can be seen as well in Figure
\ref{fig:MCData:25GeV:trackLengthFirstLayer} showing the primary tracks. All
physics lists recreate the peak around the full detector length of
$38\;$layers. This indicates a good understanding of the simulation of the
beamline, including the decay of pions to muons and the trigger. The exponential
fall is interrupted for track lengths around 28-30 layers. This is due to the
change in the geometry of the AHCal from fine to coarse after the first 30
layers. Two particles resolved in the fine granularity might become unresolved
and give non-isolated hits, therefore ending the track. Hence many tracks will
stop in layer 30, leading to the observed length distribution for tracks starting
at the front face of the AHCal. This discontinuity in track length can be seen both for
testbeam data as well as for all simulations. However, the discontinuity is more pronounced in
all considered physics lists compared to testbeam data, indicating potential
problems in the digitization, i.e. in the simulation of the optical crosstalk
which might differ for different tile sizes and hence creates a different number
of fake hits, changing the number of isolated hits.

\section{Conclusion}
\label{sec:conclusion}
A simple tracking algorithm has been developed that is capable of identifying
tracks created by minimum ionizing particles in hadronic showers. The algorithm
relies on isolated hits and works on a layer-by-layer basis. The intrinsic track
properties track angle and track length are used as
parameters in a comparison between testbeam data and simulations created with
various physics lists. For the given data the four physics lists \qgspBert,
\qgspBertTrv, \ftfBic~and~\ftfpBert~all give results that are close together
and comparable to testbeam data, with a slight advantage in favor of the
\physicsList{QGSP\_BERT(\_TRV)} lists.


\begin{footnotesize}


\end{footnotesize}


\end{document}